\input{aipcheck}

\documentclass[
    final            
  ]
  {aipproc}

\layoutstyle{6x9}

\usepackage[]{natbib}

\def\aj{AJ}

\def\apj{ApJ}

\def\apjs{ApJS}

\def\aap{A\&A}

\def\pasp{{PASP}}

\begin{document}

\title{Rotation and Activity in Late-type M Dwarfs}


\author{Andrew A. West}{
  address={MIT Kavli Institute for Astrophysics and Space Research, 
77 Massachusetts Avenue, Cambridge, MA 02139}
, altaddress = {Astronomy Department, University of California, 601
  Campbell Hall, Berkeley, CA, 94720-3411}
}

\author{Gibor Basri}{
  address={Astronomy Department, University of California, 601
  Campbell Hall, Berkeley, CA, 94720-3411}
}

\begin{abstract}
We have examined the relationship between rotation and activity in
  14 late-type (M6-M7) M dwarfs, using high resolution spectra taken
  at the Keck Observatory and flux-calibrated spectra from the
  Sloan Digital Sky Survey. Most are inactive at a spectral type where
  H$\alpha$ emission has previously seen to be very common. We used
  the cross-correlation technique to quantify the rotational
  broadening; six of the stars in our sample have $v$sin$i$$\geq$3.5
  kms$^{-1}$.  Three of these stars do not exhibit H$\alpha$ emission,
  despite rotating at velocities where previous work has observed
  strong levels of magnetic field and stellar activity.  Our results
  suggest that rotation and activity in late-type M dwarfs may not
  always be linked, and open several additional possibilities including
  a rotation dependant activity threshold, or a Maunder-minimum
  phenomenon in fully convective stars.
\end{abstract}

\maketitle


\section{Introduction}

Many M dwarfs, which are the most abundant stars in the Milky Way,
have strong magnetic dynamos that give rise to chromospheric and
coronal heating, producing emission from the x-ray to the radio.
Although this magnetic heating (or activity) has been observed for
decades, the exact mechanisms that control magnetic activity in M
dwarfs are still not well-understood.

In the Sun, activity is strongly linked to rotation.  The rotation in
solar type stars slows with time due to angular momentum loss from
magnetized stellar winds; as a result, magnetic activity decreases.
There is strong evidence that the rotation-activity relation extends
from stars more massive than the Sun to smaller dwarfs
\citep{Pizzolato2003, Kiraga2007}.  However, at a spectral type of
$\sim$M3 \citep[0.35 M$_{\odot}$;][]{NLDS, Chabrier1997}, stars become
fully convective.  This transition marks an important change in the
stellar interior that has been thought to affect the production and
storage of internal magnetic fields.

A few studies have uncovered evidence of a possible rotation-activity
(using H$\alpha$) relation extending past the M3 convective transition
and into the brown dwarf regime \citep{D98, MB03, RB07}.  However, the
lack of an unbiased sample of high resolution spectra of late-type M
dwarfs complicates the situation.

Using over 30,000 spectra from the Sloan Digital Sky Survey
\citep[SDSS;][]{ DR6}, \citet{W06, W08} showed that the activity
fraction of M dwarfs varies as a function of stellar age (using
Galactic height as a proxy for age) and that the H$\alpha$ activity
lifetime for M6-M7.5 stars is 7-8 Gyr.  Nearby samples of late-type M
dwarfs are therefore biased towards young populations with high levels
of activity; until recently every known M7 dwarf was observed to be
magnetically active \citep{Hawley96, Gizis2000, W04}.

We present results from our study of the $v$sin$i$ rotation velocities
for a small sample of M6-M7 dwarfs, most of which were selected to be
inactive or weakly active from the SDSS low-mass star spectroscopic
sample.



\section{Data}

Our sample was selected from the \citet{W08} Sloan Digital Sky
Survey (SDSS) M dwarf catalog, a spectroscopic sample of almost 40,000
M and L-type dwarfs.  We selected the brightest M7 and M6 stars which
were either inactive or weakly active (as measured by their H$\alpha$
emission).  12 stars were selected using these criteria.  Two
additional active M7 dwarfs were added to the sample for comparison to
previous studies.  While our sample is not a complete unbiased sample,
representative of the underlying M dwarf population, it does consist
of late-type M dwarfs with activity properties selectively different 
than previously observed.

\section{Analysis}

To measure the $v$sin$i$ rotation velocities for our sample, we used a
cross-correlation technique similar to that of previous studies
\citep[e.g.][]{D98, MB03}: we
cross-correlated each program spectrum with the spectrum of a slowly
rotating comparison star. The width of the resulting cross correlation
function is a direct probe of the rotational broadening.

To measure the rotational broadening of each spectrum, we compared the
resulting cross-correlation function to that of a rotationally
broadened template. The GL 406 template was rotationally broadened to
larger rotation velocities using the technique of \citet{Gray92} and
cross-correlated with the original (unbroadened) template.  A $v$sin$i$ was
determined based on the best fit spun-up template.  Figure 1 shows the
cross-correlation function of SDSS094738.45+371016.5 with GL406 in the
7080-7140\AA~ region (solid) compared with the cross-correlation
function of GL406 with the best-fit rotationally broadened GL406
spectrum (dotted; 6 kms$^{-1}$), and the auto-correlation function of
GL406 (dashed; 0 kms$^{-1}$ broadening).  The cross-correlation
reveals that SDSS094738.45+371016.5 appears to be rotating with a
velocity $\geq$ 6 kms$^{-1}$

All of the spectra were spectral typed by eye using the Hammer
spectral analysis package \citep{Covey07} on the SDSS spectra.  We
measured the equivalent widths (EW) of the H$\alpha$ emission lines in
both the SDSS and Keck spectra.  The Keck spectra are more sensitive
to low levels of emission (they can be distinguished better against
the pervasive molecular features); it is also true in general that
equivalent widths tend to be smaller when measured from high
resolution spectra.  Almost all of our targets chosen to be inactive
at low resolution proved inactive even at high resolution, and the EWs
when detected were similar. 

\section{Results}

Six of the fourteen M6-M7 dwarfs in our sample have detectable
rotation. 3 of the rotating stars have measurable activity but the
other 3 show no signs of activity in any of the emission lines in
either the SDSS or Keck spectra.  The cross-correlation shown in
Figure 1 (SDSS094738.45+371016.5) is an example of one of the inactive
M7 dwarfs that appears to be rotating despite not being magnetically
active.

\begin{figure}[!htbp]
\centering
\includegraphics[scale=0.4]{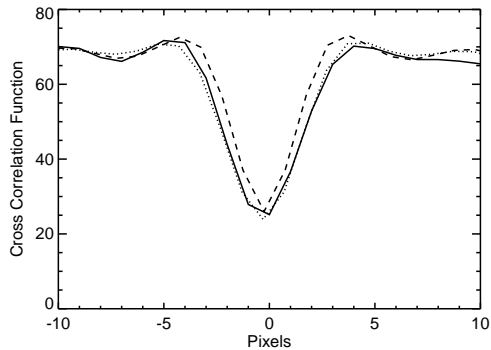} 
\caption{Cross correlation function of SDSS094738.45+371016.5 with
  GL406 in the 7080-7140 \AA~ region (solid) compared with the
  cross-correlation function of GL406 with the best-fit rotationally
  broadened GL406 spectrum (dotted; 6 kms$^{-1}$), and the
  auto-correlation function of GL406 (dashed; 0 kms$^{-1}$
  broadening). The cross-correlation reveals that
  SDSS094738.45+371016.5 appears to be rotating with a velocity $\geq$
  6 kms$^{-1}$ despite not having any signs of activity in either the
  SDSS or Keck spectra.}
\label{cc}
\end{figure}

Figure 2 shows L$_{\rm{H}\alpha}$/L$_{bol}$ (activity) as a function
of $v$sin$i$ for the M6-M7.5 dwarfs from this paper.  Lower limits in
both velocity and activity denote the levels to which our sample could
probe.  All previous M6-M7.5 dwarfs were found to be active, while 9
of the 14 stars in our sample show no activity in either the SDSS or
Keck spectra.  The lack of activity is not surprising since that was
our main selection criterion, however it highlights the fact that we
are probing a sample with very different properties than previously
studied.

\begin{figure}[!htbp]
\centering
\includegraphics[scale=0.5]{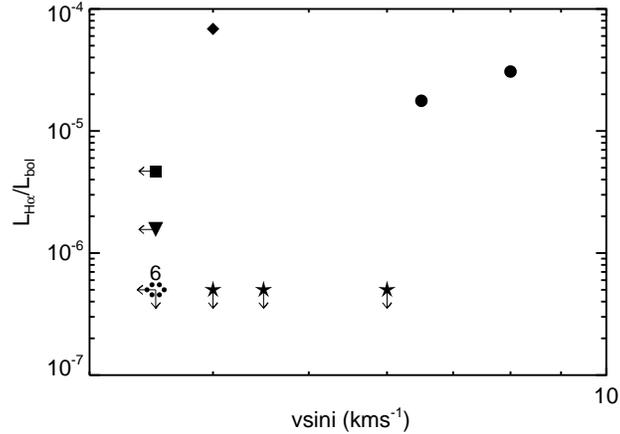} 
\caption{L$_{\rm{H}\alpha}$/L$_{bol}$ (activity) as a function of
  $v$sin$i$ (rotation) for the M6-M7.5 dwarfs from this study. 
  L$_{\rm{H}\alpha}$/L$_{bol}$ values were calculated from Equivalent
  Width measurements using the $\chi$ conversions of \citet{WHW04}.  Our sample includes M6-M7 dwarfs with both measured
  rotation as well as activity from the SDSS and Keck spectra (filled
  circles), measured rotation and activity from the Keck spectra (CaII
  K detected in SDSS; filled diamonds), activity from the Keck spectra
  but no rotation (filled triangles), activity from the SDSS spectra
  but no rotation (filled squares), no rotation or activity
  (hexagonally alligned dots; number denotes number of stars) and
  measured rotation but no activity (filled stars). Lower limits in
  both velocity and activity denote the levels to which our sample
  (and previous studies) could probe.  All previous M6-M7.5 dwarfs
  were found to be active, while 9 of the 14 stars in our sample show
  no activity in either the SDSS or Keck spectra.  The dearth of
  inactive late-type M dwarfs in previous studies is due to a
  selection effect that biases nearby samples to younger, more active
  stars \citep{W06, W08}. 3 of our stars show strong evidence
  for rotation despite having no activity.}
\label{vsini}
\end{figure}

\section{Discussion}

We conducted high resolution spectral observations of 14 M6-M7 dwarfs
and found $v$sin$i$ rotation velocities for 6 for the stars.  Three of
the stars showed both activity and rotation, 6 of the stars showed
neither rotation nor activity, 2 of the stars showed activity but no
rotation and 3 stars showed rotation but no activity.  These results
are in contrast with previous studies that found a strong connection
between rotation and activity in all (active) M6-M7 dwarfs \citep{D98, MB03, RB07}.  Our
sample is the first rotation study to include M6-M7 dwarfs that are
inactive.

\begin{theacknowledgments}
  We gratefully acknowledge the support of the AAS International
  Travel Grant.
\end{theacknowledgments}





\IfFileExists{\jobname.bbl}{}
 {\typeout{}
  \typeout{******************************************}
  \typeout{** Please run "bibtex \jobname" to optain}
  \typeout{** the bibliography and then re-run LaTeX}
  \typeout{** twice to fix the references!}
  \typeout{******************************************}
  \typeout{}
 }

\end{document}